\documentclass[aps,prd,amssymb,eqsecnum,showpacs,nofootinbib,floatfix,twocolumn,a4paper]{revtex4-1}

\usepackage{amsmath}
\usepackage{graphicx}

\allowdisplaybreaks

\newcommand{\be}{\begin{equation}}

\newcommand{\ee}{\end{equation}}

\newcommand{\md}{\mathrm{d}}

\newcommand{\pp}{(p^2)}
\newcommand{\ppn}{p^2}
\newcommand{\np}{(np)}

\newcommand{\bea}{\begin{eqnarray}}
\newcommand{\eea}{\end{eqnarray}}
\newcommand{\ben}{\begin{eqnarray*}}
\newcommand{\een}{\end{eqnarray*}}

\def\pitt#1#2{\pi^{#1#2}_{\rm TT}}
\def\pittv#1#2{\pi^{#1#2}_{\rm TT}}

\def\htt#1#2{h^{\rm TT}_{#1#2}}
\def\httiv#1#2{h^{\rm TT}_{#1#2}}

\def\httivdot#1#2{\dot{h}^{\rm TT}_{#1#2}}

\def\ghtt#1#2#3{h^{\rm TT}_{#1#2,#3}}

\begin{document}

\title{Dimensional regularization of local singularities
in the 4th post-Newtonian two-point-mass Hamiltonian}

\author{Piotr Jaranowski}
\email{pio@alpha.uwb.edu.pl}
\affiliation{Faculty of Physics,
University of Bia{\l}ystok,
Lipowa 41, 15--424 Bia{\l}ystok, Poland}

\author{Gerhard Sch\"afer}
\email{gos@tpi.uni-jena.de}
\affiliation{Theoretisch-Physikalisches Institut,
Friedrich-Schiller-Universit\"at,
Max-Wien-Pl.\ 1, 07743 Jena, Germany}

\date{}

\begin{abstract}

The article delivers the only still unknown coefficient in the 4th 
post-Newtonian energy expression for binary point masses on circular orbits 
as function of orbital angular frequency. Apart from a single coefficient, 
which is known solely numerically, all the coefficients are given as exact 
numbers. The shown Hamiltonian is presented in the center-of-mass frame and 
out of its 57 coefficients 51 are given fully explicitly. Those coefficients 
are six coefficients more than previously achieved
[Jaranowski/Sch\"afer, Phys.\ Rev.\ D \textbf{86}, 061503(R) (2012)].
The local divergences in the point-mass model 
are uniquely controlled by the method of dimensional regularization.
As application, the last stable circular orbit is determined
as function of the symmetric-mass-ratio parameter.   

\end{abstract}

\pacs{04.25.Nx, 04.30.Db, 97.60.Jd, 97.60.Lf}

\maketitle

\section{Introduction}

The measurement of gravitational waves will finally reveal the motion of 
binary black holes with highest precision including the 4th post-Newtonian 
(4PN) approximation under consideration in the present article \cite{LIGO2011}. 
Particularly important is the knowledge of the last stable circular orbit 
(LSCO) beyond which binary systems do plunge and undergo merging.
Though PN schemes are not quite appropriate for describing plunge orbits
reasonable estimates can be made for the LSCO.
For achieving more, the effective-one-body (EOB) approach by Damour and collaborators \cite{Damour2012},
which is strongly based on PN approximations, allows treatment of the dynamics beyond LSCO
(see, e.g., \cite{BDGNR2011,BBLT2012}).
Of most urgent need is the knowledge of the PN dynamics beyond 3.5PN order, particularly the 4PN order. 

In this communication we will report on the calculation of six more 
coefficients of the 4PN center-of-mass Hamiltonian for binary point-mass systems 
which resulted in the knowledge of 51 coefficients out of total 57 ones. 
Without knowledge of the missing six coefficients, the full energy expression 
of binary systems on circular orbits is derived for the first time taking 
into account a coefficient which was recently calculated numerically \cite{TBW2012}.
The numerically-only-known coefficient is uniquely connected with the global structure of the near zone 
and needs for its calculation the regularization of infrared divergences.
All the other coefficients in the energy expression for circular orbits at 4PN order --
apart from the intimately connected with logarithmic function term linear in the ratio of reduced mass to total mass --
are calculable with the aid of the technic of dimensional regularization,
in this way controlling the ultraviolet divergences resulting from the applied point-mass model.
It is worth noticing that the corresponding logarithmic terms in the Hamiltonian presented below
were uniquely calculated with the aid of analytic regularization in three-dimensional space;
here, pure dimensional regularization would have supplied logarithmic expressions not in agreement with the Poincar\'e algebra.
The previous achievements in the expression for the energy on circular orbits at 4PN \cite{JS2012}
are confirmed through \cite{D2010,BDTW2010} and, most recently, by \cite{FS2012}. 

We employ the following notation:
$\mathbf{x}=\left(x^i\right)$ ($i=1,2,3$) denotes a point in the 3-dimensional
Euclidean space $\mathbb{R}^3$ endowed with a standard Euclidean metric
and a scalar product (denoted by a dot).
Letters $a$ and $b$ ($a,b=1,2$) are body labels,
so $\mathbf{x}_a\in\mathbb{R}^3$ denotes the position of the $a$th point mass.
We also define  ${\bf r}_a\equiv\mathbf{x}-\mathbf{x}_a$, $r_a \equiv |{\bf r}_a|$,
${\bf n}_a\equiv{\bf r}_a/r_a$; and for $a\ne b$, 
${\bf r}_{ab}\equiv\mathbf{x}_a-\mathbf{x}_b$,
$r_{ab} \equiv |{\bf r}_{ab}|$,
${\bf n}_{ab} \equiv {\bf r}_{ab}/r_{ab}$;
$|\cdot|$ stands here for the Euclidean length of a vector. 
The linear momentum vector of the $a$th body is denoted by $\mathbf{p}_a=\left(p_{ai}\right)$,
and $m_a$ denotes its mass parameter.
We abbreviate $\delta\left({\bf x}-{\bf x}_a\right)$ by $\delta_a$.
Extensive use has been made of the computer-algebra system \textsc{Mathematica}.

\section{The ADM canonical approach in $d$ space dimensions}

Let $D\equiv{d+1}$ denote the (analytically continued) space-time dimension.
The ADM approach \cite{ADM62} uses a $d+1$ split of the coupled gravity-matter
dynamics and works with the canonical pairs $(x_a^i,p_{ai})$ and $(g_{ij},\pi^{ij})$
($i,j,k,\ldots$ denote spatial indices taking formally $d$ values; $a,b=1,2$ labels the particles).
The dimensionally continued hamiltonian and momentum constraints read
(in units where $16\pi\,G_D=c=1$ with $G_D$ denoting
the generalized Newtonian gravitational constant and $c$ the speed of light)
\begin{subequations}
\begin{align}
\sqrt{g}\,R &= \frac{1}{\sqrt g} \left(g_{ik} \, g_{j\ell} \, \pi^{ij} \, 
\pi^{k\ell} - \frac{1}{d-1} \ (g_{ij} \, \pi^{ij})^2 \right)
\nonumber\\[1ex]&\quad
+ \sum_a (m_a^2 + g_a^{ij} \, p_{ai} \, p_{aj})^{\frac{1}{2}} \, \delta_a,
\\[2ex]
-2\,D_j\,\pi^{ij} &= \sum_a g_a^{ij} \, p_{aj} \, \delta_a. 
\end{align}
\end{subequations}
Here $R$ denotes the space curvature of the hypersurface $t=\text{const}$,
$g$ is the determinant of the matrix $(g_{ij})$,
$\delta_a$ denots $d$-dimensional Dirac delta distribution
(with $\int{\md^dx\,\delta_a}=1$), 
$g_a^{ij}\equiv{g^{ij}({\bf x}_a)}$ is the finite part of metric
evaluated at the particle position (which can be perturbatively 
unambiguously defined), and $D_j$ is the $d$-dimensional 
covariant derivative (acting on a tensor density of weight one).

Taking into account the appropriate coordinate conditions
\be
g_{ij} = \Psi^{\frac{4}{d-2}} \delta_{ij} + {\htt ij},
\quad \pi^{ii}=0,
\ee
where 
\be
\displaystyle\Psi = 1+ \frac{1}{4}\frac{d-2}{d-1}\phi
\ee
holds and where ${\htt ij}$ is a symmetric transverse-traceless (TT) quantity, ${\htt ii}={\ghtt ijj}=0$,
and the field momentum $\pi^{ij}$ is splitted into its longitudinal and TT parts, 
respectively $\pi^{ij} = \tilde{\pi}^{ij} + {\pitt ij}$,
with $\tilde{\pi}^{ij} = \partial_i\pi^j +\partial_j\pi^i - \frac{2}{d}\delta^{ij}\partial_k\pi^k$,   
from the constraint equations, the reduced Hamiltonian results in the form,
\be
H_\text{red}\big[\mathbf{x}_a,\mathbf{p}_a,{\htt ij},{\pitt ij}\big]
= -\int \md^dx\,\Delta\phi\big[\mathbf{x}_a,\mathbf{p}_a,{\htt ij},{\pitt ij}\big].
\ee
This Hamiltonian describes the evolution of the matter and independent gravitational field variables.

An autonomous conservative Hamiltonian can be obtained 
through the transition to a Routhian description,
\be
R\big[\mathbf{x}_a,\mathbf{p}_a,{\httiv ij},{\httivdot ij}\big]
\equiv H_\text{red}  - \int\md^dx\,{\pittv ij}{\httivdot ij}.
\ee
Then the matter Hamiltonian reads
\be
H(\mathbf{x}_a,\mathbf{p}_a)
= R\big[{\bf x}_a,{\bf p}_{a},
{\httiv ij}(\mathbf{x}_a,\mathbf{p}_a),
{\httivdot ij}(\mathbf{x}_a,\mathbf{p}_a)\big],
\ee 
where higher-order time derivatives are eliminated through lower-order equations of motion \cite{DS1991},
corresponding to a canonical transformation.

\section{Dimensional regularization of local divergences}

Dimensional regularization (DR) has shown its power in the previous 3PN calculations
where unique results were achieved with using Dirac delta distributions
as source functions \cite{DJS2001,BDEF04,FS2011a} (also see \cite{DJS2008,Chu2009}).
Distribitional derivatives must still be applied
but crucial for the success of the DR technic
is the preservation of the Leibniz rule for differentiations
and the related treatment of the finite part of products of singular functions.
On the other side, at the 3PN Hamiltonian level, all occuring $1/(d-3)$-poles
together with the connected logarithmic terms do cancel each other \cite{DJS2001}.
As efficient DR has proven in controlling local divergences,
it failed in the treatment of long-range divergences.
Those infrared divergences have nothing to do with the used point-mass model
but rely on the very definition of the near zone.
Within our formalism, those divergences are still not fully under control and need futher investigations.
Only the logarithmic terms resulting in this context we were able to calculate uniquely.

To compute the 4PN Hamiltonian $H_\text{4PN}$ we have splitted
the Hamiltonian and Hamiltonian density $h_\text{4PN}$ into two parts,
\begin{subequations}
\begin{align}
H_\text{4PN}(d) &= \int \md^dx \, h_\text{4PN}(d)
= H_\text{4PN}^\text{loc}(d) + H_\text{4PN}^\text{inf}(d),
\\
H_\text{4PN}^\text{loc}(d) &\equiv \int \md^d x \,h_\text{4PN}^\text{loc}(d),
\\
H_\text{4PN}^\text{inf}(d) &\equiv \int \md^d x \,h_\text{4PN}^\text{inf}(d),
\end{align}
\end{subequations}
where the integral over $h_\text{4PN}^\text{loc}$ is convergent at spatial infinity albeit it is locally divergent,
and the integral over $h_\text{4PN}^\text{inf}$ is divergent at spatial infinity
but $h_\text{4PN}^\text{inf}$ is locally integrable.
In the present paper we have computed the 3-dimensional limit of the first integral.
To regularize local divergencies we have employed the DR technic
in a way described in detail in Ref.\ \cite{DJS2001}.
It means that in fact we have computed the difference
\be
\Delta H_\text{4PN} \equiv \lim_{d\to3} H_\text{4PN}^\text{loc}(d) - H_\text{4PN}^\text{RH loc}(3),
\ee
where $H_\text{4PN}^\text{RH loc}(3)$ is the ``local part'' of the Hamiltonian
obtained by means of the 3-dimensional Riesz-implemented Hadamard (RH) regularization
defined in Refs.\ \cite{J1997,JS1998,DJS2000b}.
The way of computing this difference was devised in Secs.\ 3 and 4 of \cite{DJS2001}
and can be summarized as follows.
Let us consider some global
(i.e.\ performed over the whole ${\mathbb R}^3$ space) integral, which develops only local poles.
Let us denote its integrand by $i(\mathbf{x})$. Then the value of the integral,
after performing the RH regularization in 3 dimensions,
usually has the structure
\begin{align}
I^{\text{RH}}&(3;\varepsilon_1,\varepsilon_2) =
\int_{{\mathbb R}^3} i({\mathbf x})
\Big(\frac{r_1}{s_1}\Big)^{\varepsilon_1} \Big(\frac{r_2}{s_2}\Big)^{\varepsilon_2} \, \md^3{\mathbf x}
\nonumber\\
&= A + c_1(3) \Big(\frac{1}{\varepsilon_1} + \ln\frac{r_{12}}{s_1} \Big)
+ c_2(3) \Big(\frac{1}{\varepsilon_2} + \ln\frac{r_{12}}{s_2} \Big)
\nonumber\\
&\quad + \mathcal{O}(\varepsilon_1,\varepsilon_2).
\end{align}
Here $s_1$ and $s_2$ are arbitrary 3-dimensional regularization scales.
To find the DR correction to the integral $I^{\text{RH}}(3;\varepsilon_1,\varepsilon_2)$,
related with the local poles at, say, $\mathbf{x}=\mathbf{x}_1$,
it is enough to consider this part of the integrand $i(\mathbf{x})$ which develops
logarithmic singularities, i.e.\ which locally behaves like $1/r_1^3$,
and it is enough to consider the integral of this part 
over the ball $B(\mathbf{x}_1,{\ell_1})$ of radius $\ell_1$ surrounding the particle $\mathbf{x}_1$.
The RH regularized value of this integral reads
\begin{align}
I_1^{\text{RH}}(3;\varepsilon_1) &\equiv
\int_{B(\mathbf{x}_1,{\ell_1})} \bar{c}_1(3;\mathbf{n}_1) \, r_1^{-3}
\Big(\frac{r_1}{s_1}\Big)^{\varepsilon_1} \, \md^3 {\mathbf r}_1
\nonumber\\
&= c_1(3) \int_0^{\ell_1} r_1^{-3} \Big(\frac{r_1}{s_1}\Big)^{\varepsilon_1} \, r_1^2 \, \md r_1,
\end{align}
where $c_1(3)$ is the angle-averaged value of the coefficient $\bar{c}_1(3;\mathbf{n}_1)$.
The expansion of the integral $I_1^{\text{RH}}(3;\varepsilon_1)$ around $\varepsilon_1=0$ equals
\be
I_1^{\text{RH}}(3;\varepsilon_1) = \frac{c_1(3)}{\varepsilon_1}
+ c_1(3) \ln \frac{\ell_1}{s_1} + \mathcal{O}(\varepsilon_1).
\ee
In the next step one computes the $d$-dimensional version
of the integral $I_1^{\text{RH}}(3;\varepsilon_1)$.
Let us call it $I_1(d)$. It has the structure
\begin{align}
\label{I1def}
I_1(d) &\equiv \ell_0^{k(d-3)} \int_{B(\mathbf{x}_1,{\ell_1})}
\bar{c}_1(d;{\mathbf n}_1) \, r_1^{6-3d}\, \md^d {\mathbf r}_1
\nonumber\\
&= \ell_0^{k(d-3)} c_1(d) \int_0^{\ell_1} r_1^{6-3d} \, r_1^{d-1}\, \md r_1,
\end{align}
where $\ell_0$ is the scale which relates the Newtonian gravitational constant $G_N$
with the $D$-dimensional gravitational constant $G_D$,
$$
G_D = G_N\,\ell_0^{d-3},
$$
and the number $k$ indicates the momentum-order of the term
[the term with $k$ is of the order of $\mathcal{O}(p^{10-2k})$,
where $k=1,\ldots,5$].
The radial integral in Eq.\ \eqref{I1def} is convergent
if the real part of $d$ fulfills $\Re(d)<3$.
The expansion of the integral $I_1(d)$ around $\varepsilon\equiv d-3=0$ reads
\be
I_1(d) = -\frac{c_1(3)}{2\varepsilon} -\frac{1}{2} c'_1(3)
+ c_1(3) \ln\frac{\ell_1}{\ell_0} + \mathcal{O}(\varepsilon).
\ee
Let us observe that the coefficient $c'_1(3)$ usually depends on $\ln r_{12}$,
so it has the structure
\be
c'_1(3) = c'_{11}(3) + c'_{12}(3) \ln\frac{r_{12}}{\ell_0}.
\ee
Therefore the DR correction will also change the terms $\propto\ln{r_{12}}$.

The DR correction of the integral
$I^{\text{RH}}(3;\varepsilon_1,\varepsilon_2)$ relies on replacing this integral by
\be
I^{\text{RH}}(3;\varepsilon_1,\varepsilon_2) + \Delta I_1 + \Delta I_2,
\ee
where
\be
\Delta I_a \equiv \lim_{d\to3} I_a(d)
- \lim_{\varepsilon_1\to3} I_1^{\text{RH}}(3;\varepsilon_1),
\quad a=1,2.
\ee
The corrected value of $I^{\text{RH}}(3;\varepsilon_1,\varepsilon_2)$ thus reads
\begin{align}
\label{IRHcorr}
&I^{\text{RH}}(3;\varepsilon_1,\varepsilon_2) + \Delta I_1 + \Delta I_2
= A -\frac{c_1(3)+c_2(3)}{2\varepsilon}
\nonumber\\&\quad
- \frac{1}{2} \big(c'_1(3) + c'_2(3)\big)
+ \big(c_1(3) + c_2(3)\big) \ln\frac{r_{12}}{\ell_0}
\nonumber\\ &=
A -\frac{c_1(3)+c_2(3)}{2\varepsilon}
- \frac{1}{2} \big(c'_{11}(3) + c'_{21}(3)\big)
\nonumber\\&
+ \big(c_1(3) - \frac{1}{2}c'_{12}(3) + c_2(3)
- \frac{1}{2}c'_{22}(3)\big) \ln\frac{r_{12}}{\ell_0}.
\end{align}
Note that all poles $\propto1/\varepsilon_1,1/\varepsilon_2$
and all terms depending on $\ln\ell_1$, $\ln\ell_2$ or $\ln s_1$, $\ln s_2$ cancel each other.
The result \eqref{IRHcorr} is as if all computations were fully done in $d$ dimensions.

\section{Center-of-mass Hamiltonian}

Making use of the procedure described in the previous section we have computed DR corrections
to all logarithmically divergent terms contributing to $H_\text{4PN}^\text{loc}(3)$.
After summing up all these corrections we have obtained terms proportional to $1/\varepsilon$
as well as to $\ln(r_{12}/\ell_0)$. We have shown that both kind of terms can be removed from the Hamiltonian
by adding a total time derivative. This way we have obtained free of poles and logarithmic terms
local part $H_\text{4PN}^\text{loc}(3)$ of the 3-dimensional 4PN Hamiltonian.
We will show here the explicit form of this part in the center-of-mass reference frame.
Let us note that the DR corrections are needed only for some terms of the order 4, 2, and 0 in momenta.
All terms of the order 10, 8, and 6 were calculated in \cite{JS2012}
by means of 3-dimensional RH regularization.

The center-of-mass frame is defined by the condition ${\bf p}_1+{\bf p}_2=0$.
It is convenient to use the following reduced variables:
$\mathbf{r}\equiv\mathbf{r}_{12}/(GM)$ (with $r\equiv|\mathbf{r}|$ and $\mathbf{n}\equiv\mathbf{r}/r$),
$\mathbf{p}\equiv\mathbf{p}_1/\mu$, where $M \equiv m_1 + m_2$ is the total mass of the system
and $\mu \equiv m_1m_2/M$ is its reduced mass.
We also introduce the reduced Hamiltonian $\hat{H}\equiv(H-Mc^2)/\mu$
which depends on masses only through the symmetric mas ratio $\nu\equiv\mu/M$
($0\le\nu\le1/4$; $\nu=0$ is the test-mass limit and $\nu = 1/4$ holds for equal masses).

We have checked that the noncomputed part of the Hamiltonian, $H_\text{4PN}^\text{inf}(3)$,
in the center-of-mass frame and in the reduced form
$\hat{H}_\text{4PN}^\text{inf}(3)\equiv H_\text{4PN}^\text{inf}(3)/\mu$, is proportional to $\nu$.
It contains logarithmic terms which result
from divergences of the instantaneous near-zone metric when going to large distances.
They were computed in Ref.\ \cite{JS2012} (see also Refs.\ \cite{D2010,BDTW2010}).
In the logarithmic terms $\hat{s}$ is a regularization scale
[$\hat{s}\equiv{s/(GM)}$, where $s$ is a regularization scale with dimension of length].

The 4PN-accurate conservative Hamiltonian $\hat{H}(\mathbf{r},\mathbf{p})$
in the center-of-mass frame equals
\begin{align}
\hat{H}(\mathbf{r},\mathbf{p}) &= \hat{H}_\text{N}(\mathbf{r},\mathbf{p})
+ \hat{H}_\text{1PN}(\mathbf{r},\mathbf{p}) + \hat{H}_\text{2PN}(\mathbf{r},\mathbf{p})
\nonumber\\[1ex]
&\quad + \hat{H}_\text{3PN}(\mathbf{r},\mathbf{p}) + \hat{H}_\text{4PN}(\mathbf{r},\mathbf{p}).
\end{align}
The reduced Hamiltonians from $\hat{H}_\text{N}$ to $\hat{H}_\text{3PN}$
can be found in Eq.\ (3.6) of \cite{DJS2000b} (where one has to put
$\omega_{\text{static}}=0$ and $\omega_{\text{kinetic}}=41/24$, see \cite{DJS2001}).
The 4PN reduced Hamiltonian $\hat{H}_\text{4PN}$ was partially computed in Ref.\ \cite{JS2012}.
For convenience of the reader we repeat here the main result of \cite{JS2012},
i.e.\ the Hamiltonian $\hat{H}_\text{4PN}$ written in the following form
[taken from Eq.\ (3.1) of \cite{JS2012};
here $\pp\equiv\mathbf{p}\cdot\mathbf{p}$ and $\np\equiv\mathbf{n}\cdot\mathbf{p}$]:
\begin{widetext}
\begin{align}
\label{hatH}
c^8\,{\hat H}_\text{4PN}(\mathbf{r},\mathbf{p}) &=
\left(
\frac{7}{256}
-\frac{63}{256}\nu
+\frac{189}{256}\nu^2
-\frac{105}{128}\nu^3
+\frac{63}{256}\nu^4
\right)\pp^5
\nonumber\\
&\quad
+ \Bigg\{
\frac{45}{128}\pp^4
-\frac{45}{16}\pp^4\nu
+\left(
\frac{423}{64}\pp^4
-\frac{3}{32}\np^2\pp^3
-\frac{9}{64}\np^4\pp^2
\right)\nu^2
\nonumber\\
&\qquad\quad
+ \left(
-\frac{1013}{256}\pp^4
+\frac{23}{64}\np^2\pp^3
+\frac{69}{128}\np^4\pp^2
-\frac{5}{64}\np^6\ppn
+\frac{35}{256}\np^8
\right)\nu^3
\nonumber\\
&\qquad\quad
+ \left(
-\frac{35}{128}\pp^4
-\frac{5}{32}\np^2\pp^3
-\frac{9}{64}\np^4\pp^2
-\frac{5}{32}\np^6\ppn
-\frac{35}{128}\np^8
\right)\nu^4
\Bigg\}\frac{1}{r}
\nonumber\\
&\quad
+ \Bigg\{
\frac{13}{8}\pp^3
+ \left(
-\frac{791}{64}\pp^3
+\frac{49}{16}\np^2\pp^2
-\frac{889}{192}\np^4\ppn
+\frac{369}{160}\np^6
\right)\nu
\nonumber\\
&\qquad\quad
+ \left(
\frac{4857}{256}\pp^3
-\frac{545}{64}\np^2\pp^2
+\frac{9475}{768}\np^4\ppn
-\frac{1151}{128}\np^6
\right)\nu^2
\nonumber\\
&\qquad\quad
+ \left(
\frac{2335}{256}\pp^3
+\frac{1135}{256}\np^2\pp^2
-\frac{1649}{768}\np^4\ppn
+\frac{10353}{1280}\np^6
\right)\nu^3
\Bigg\}\frac{1}{r^2}
\nonumber\\
&\quad
+ \Bigg\{
\frac{105}{32}\pp^2
+ \left[
C_{41}(\mathbf{r},\mathbf{p})
+ \left(
\frac{237}{40}\pp^2
-\frac{1293}{40}\np^2\ppn
+\frac{97}{4}\np^4
\right) \ln\frac{r}{\hat{s}} \right]\nu
\nonumber\\
&\qquad\quad
+ C_{42}(\mathbf{r},\mathbf{p})\,\nu^2
+ \left(
-\frac{553}{128}\pp^2
-\frac{225}{64}\np^2\ppn
-\frac{381}{128}\np^4
\right)\nu^3 
\Bigg\}\frac{1}{r^3}
\nonumber\\
&\quad
+ \Bigg\{
\frac{105}{32}\ppn
+ \left[C_{21}(\mathbf{r},\mathbf{p})
+ \left(
\frac{233}{40}\ppn
-\frac{29}{6} \np^2
\right)\ln\frac{r}{\hat{s}} \right]\nu
+ C_{22}(\mathbf{r},\mathbf{p})\,\nu^2
\Bigg\}\frac{1}{r^4}
\nonumber\\
&\quad
+ \Bigg\{
-\frac{1}{16}
+ \left[c_{01}+\frac{21}{20}\ln\frac{r}{\hat{s}}\right]\nu
+ c_{02}\,\nu^2
\Bigg\}\frac{1}{r^5}.
\end{align}
The Hamiltonian \eqref{hatH} contains the terms $C_{4i}(\mathbf{r},\mathbf{p})$, $C_{2i}(\mathbf{r},\mathbf{p})$ ($i=1,2)$
and the coefficients $c_{01}$ and $c_{02}$ which were not computed in Ref.\ \cite{JS2012}.
In the present paper we have calculated, making use of dimensional regularization,
the terms $C_{42}(\mathbf{r},\mathbf{p})$, $C_{22}(\mathbf{r},\mathbf{p})$
and the coefficient $c_{02}$. They read
\begin{subequations}
\label{coeffnew}
\begin{align}
C_{42}(\mathbf{r},\mathbf{p}) &= \left(-\frac{1189789}{28800} + \frac{18491}{16384}\pi^2\right)\pp^2
+ \left(-\frac{127}{3} - \frac{4035}{2048}\pi^2\right)\np^2\ppn
+ \left(\frac{57563}{1920} - \frac{38655}{16384}\pi^2\right)\np^4,
\\[1ex]
C_{2i}(\mathbf{r},\mathbf{p}) &= \left(\frac{672811}{19200} - \frac{158177}{49152}\pi^2\right)\pp
+ \left(-\frac{21827}{3840} + \frac{110099}{49152}\pi^2\right)\np^2,
\\[1ex]
c_{02} &= -\frac{1256}{45} + \frac{7403}{3072}\pi^2.
\end{align}
\end{subequations}
\end{widetext}
The still noncomputed is the coefficient $c_{01}$
and the terms $C_{41}(\mathbf{r},\mathbf{p})$ and $C_{22}(\mathbf{r},\mathbf{p})$ of the structure
\begin{subequations}
\begin{align}
C_{41}(\mathbf{r},\mathbf{p}) &= c_{411}\pp^2 + c_{412}\np^2\ppn + c_{413}\np^4,
\\[1ex]
C_{21}(\mathbf{r},\mathbf{p}) &= c_{211}\pp + c_{212}\np^2.
\end{align}
\end{subequations}

In Ref.\ \cite{JS2012} the Poincar\'e algebra relations were used \cite{DJS2000}
to uniquely determine the three quartic in momenta and proportional to $\nu^3$ coefficients.
In the present paper we have recomputed these coefficients directly
(i.e.\ without employig the Poincar\'e algebra).

\section{Circular orbits}

Making use of the Hamiltonian \eqref{hatH} together with the new results of Eqs.\ \eqref{coeffnew},
we have computed the energy of binary system moving along circular orbits
as a function of angular frequency $\omega$ (details of such computation performed
at the 3PN order can be found in Ref.\ \cite{DJS2000b}
and was also briefly summarized in Ref.\ \cite{JS2012}).
We have introduced the dimensionless PN parameter
$x\equiv(GM\omega/c^3)^{2/3}$
and we have assumed (following \cite{BDTW2010})
that the regularization scale $s$ is determined by the period $P$
of the motion along circular orbits, $s=cP$.
Making then use of the third Kepler's law at the Newtonian level
one shows that $\hat{s}=2\pi c^{-2}x^{-3/2}$.
The 4PN-accurate binding energy of the system can be written in the form
\begin{align}
\label{eofx}
& E(x;\nu) = -\frac{\mu c^2 x}{2} \biggl(
1 +  e_{\text{1PN}}(\nu)\,x + e_{\text{2PN}}(\nu)\,x^2
\nonumber\\[1ex]&
+ e_{\text{3PN}}(\nu)\,x^3
+ \Big(e_{\text{4PN}}(\nu)+\frac{448}{15}\nu\ln x\Big)\,x^4
+ {\cal O}\big(x^5\big) \biggr),
\end{align}
where the fractional corrections to the Newtonian energy at different PN orders read
\begin{widetext}
\begin{subequations}
\begin{align}
e_{\text{1PN}}(\nu) &= -\frac{3}{4} - \frac{1}{12}\nu,
\\[1ex]
e_{\text{2PN}}(\nu) &= -\frac{27}{8} + \frac{19}{8}\nu
- \frac{1}{24}\nu^2,
\\[1ex]
e_{\text{3PN}}(\nu) &= -\frac{675}{64} +
\left(\frac{34445}{576}-\frac{205}{96}\pi^2\right)\nu -
\frac{155}{96}\nu^2 - \frac{35}{5184}\nu^3,
\\[1ex]
e_{\text{4PN}}(\nu) &= -\frac{3969}{128}
+ 153.8803(1)\,\nu
+ \left(-\frac{498449}{3456} + \frac{3157}{576}\pi^2\right)\nu^2
+ \frac{301}{1728}\nu^3
+ \frac{77}{31104}\nu^4.
\end{align}
\end{subequations}
\end{widetext}
In the above formula, the 4PN coefficient at $\nu^2$ was computed for the first time.
The 4PN coefficient linear in $\nu$ was computed numerically in Ref.\ \cite{TBW2012}
[see also Eq.\ (3.1) in \cite{BBLT2012}].
The coefficientes at $\nu^3$ and $\nu^4$ computed for the first time in \cite{JS2012}
was recently confirmed by independent calculation made in \cite{FS2012}.

\begin{figure*}
\begin{center}
\begin{tabular}{cc}
\includegraphics[scale=0.75]{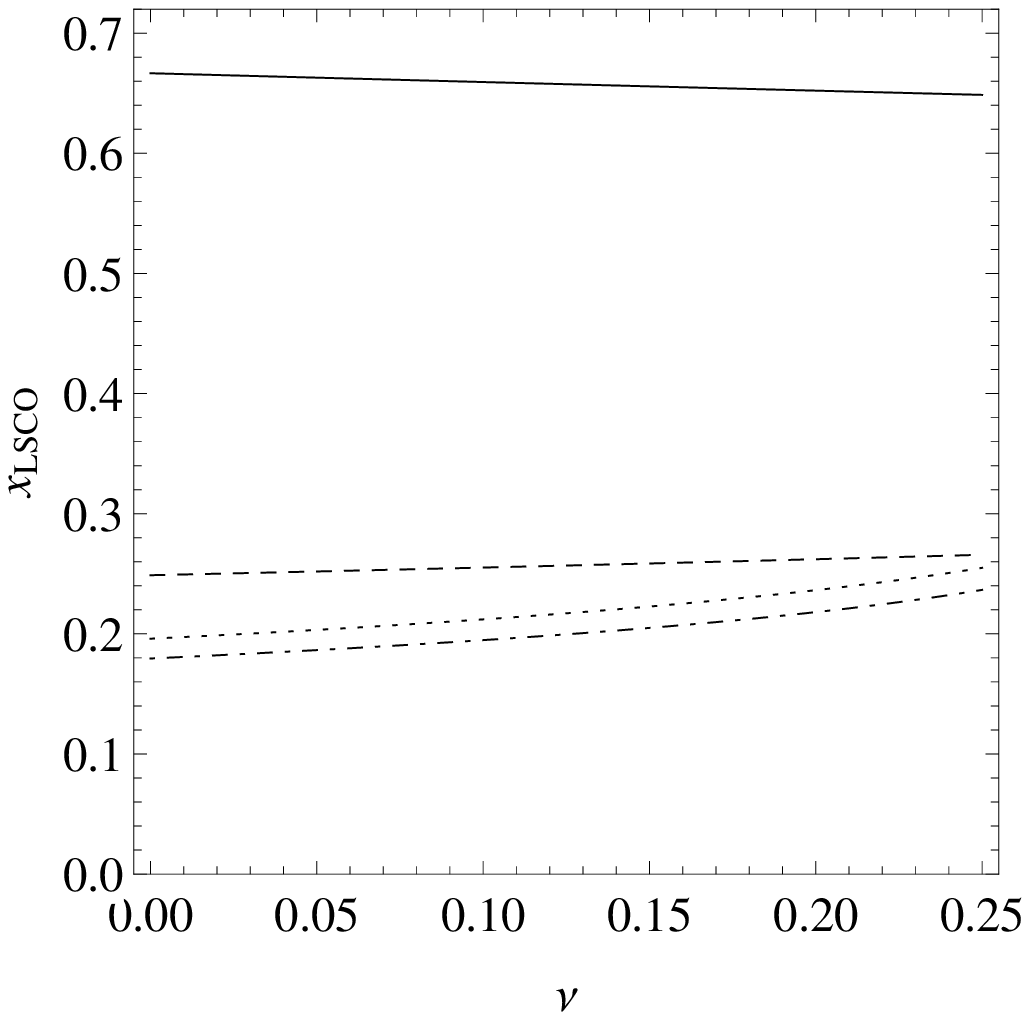}&
\includegraphics[scale=0.76]{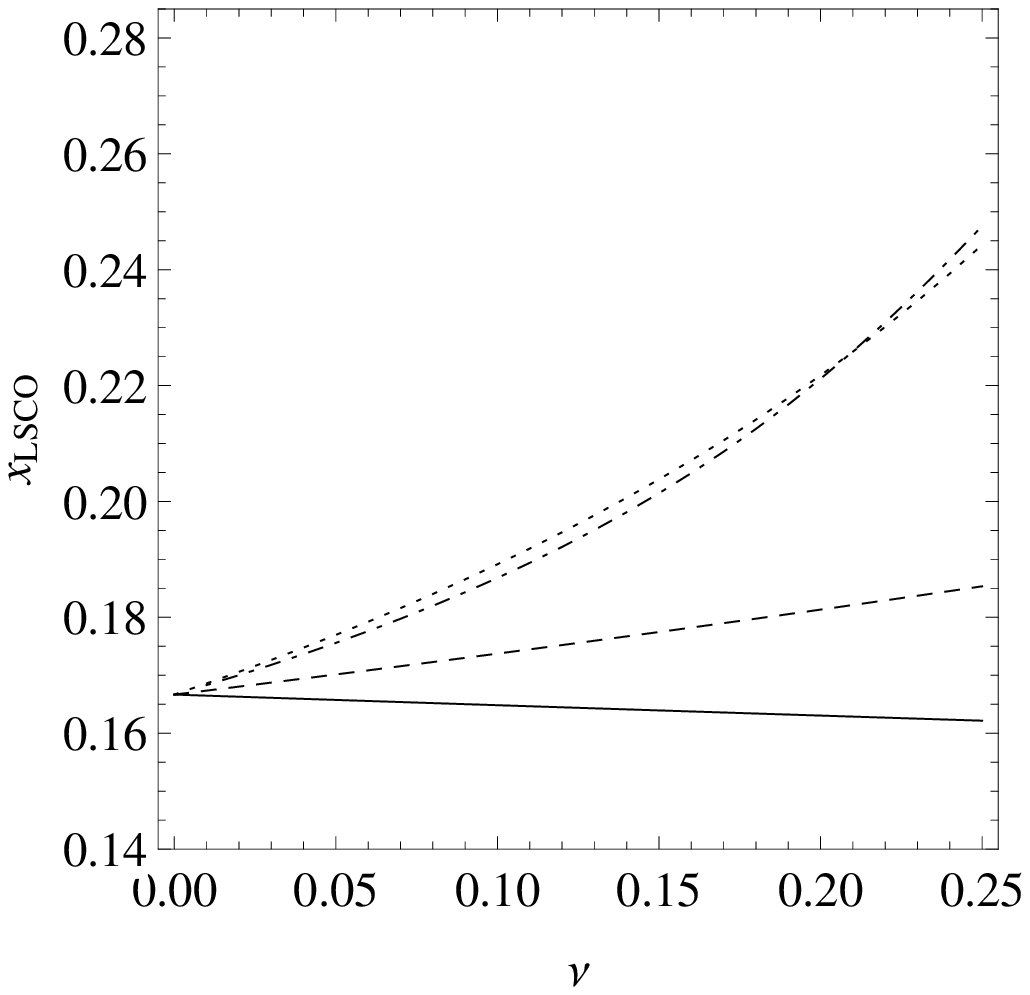}
\end{tabular}
\caption{\label{LSCO}
The location $x_\text{LSCO}$ of the LSCO
as a function of the symmetric mass ratio $\nu$
for successive PN approximations from 1PN up to 4PN.
Solid curves correspond to 1PN, dashed curves to 2PN,
dotted curves to 3PN, and dashed/dotted curves to 4PN.
Left panel: $x_\text{LSCO}$ is computed as the minimum of the function $E(x;\nu)$.
Right panel: $x_\text{LSCO}$ is obtained by means of the $j$-method.}
\end{center}
\end{figure*}

An important feature of the two-point-mass dynamics along circular orbits
is the existence of the \emph{last stable circular orbit} (LSCO) \cite{DJS2000c}.
In the test-mass limit $\nu=0$, the LSCO occurs for $x_\text{LSCO}=1/6$
what corresponds to the minimum of the function $E(x;\nu=0)$.
Therefore the most straightforward way of locating the LSCO for $\nu>0$
relies on looking for minimum of the function $E(x;\nu)$.
In Fig.\ \ref{LSCO} (left panel) we have depicted the location $x_\text{LSCO}$
of the LSCO as function of the symmetric mass ratio $\nu$
for successive PN approximations from 1PN up to 4PN.
For a given value of $\nu$, the location of the LSCO
was obtained by numerically solving the equation $\md E(x;\nu)/\md x=0$.
Let us note that the 4PN prediction for the location of the LSCO
in the test-mass limit is 0.179467, $\sim$7.7\% larger than the exact result.
We have obtained the following locations of the LSCO in the equal-mass case ($\nu=1/4$)
for the approximations from 1PN up to 4PN:
0.648649 (1PN), 0.265832 (2PN), 0.254954 (3PN), 0.236597 (4PN).

We have also found the location of the LSCO by means of more refined method,
namely the ``$j$-method'' introduced and used at the 3PN level in Ref.\ \cite{DJS2000c}.
In this method the $x_\text{LSCO}$ is defined as the minimum of the function $j^2(x)$,
where $j\equiv\mathcal{J}/(Gm_1m_2)$ is the reduced angular momentum of the system
($\mathcal{J}$ is the total angular momentum). Moreover, motivated by the form
of the function $j^2(x)$ in the test-mass limit, $j^2(x)=1/(x(1-3x))$,
Pad\'e approximants are used instead of direct Taylor expansions
(all used approximants have a pole for some $x_\text{pole}(\nu)$
which is related with the test-mass ``light-ring'' orbit occuring for $x_\text{lr}=1/3$
in the sense that $x_\text{pole}(\nu)\to1/3$ when $\nu\to0$).
The 4PN-accurate function $j^2(x)$ has the symbolic structure
$(1/x)(1+x+\ldots+x^4+x^4\ln x)$.
In the $j$-method the Taylor expansion at the 1PN level
of the symbolic form $1+x$ is replaced by Pad\'e approximant of type (0,1),
at the 2PN level $1+x+x^2$ is replaced by (1,1) approximant,
at the 3PN level $1+x+x^2+x^3$ is replaced by (2,1) approximant,
and finally at the 4PN level $1+x+x^2+x^3+x^4$ is replaced by (3,1) Pad\'e approximant
[the explicit form of the (0,1), (1,1), and (2,1) approximants
can be found in Eqs.\ (4.16) of \cite{DJS2000c}].
The results are illustrated in Fig.\ \ref{LSCO} (right panel).
At all PN levels the test-mass result is recovered exactly.
We see that the curves corresponding to 3PN-accurate and 4PN-accurate calculations almost coincide.
The locations of the LSCO in the equal-mass case $\nu=1/4$
for the approximations from 1PN up to 4PN are as follows:
0.162162 (1PN), 0.185351 (2PN), 0.244276 (3PN), 0.247515 (4PN).

\begin{acknowledgments}

P.J.\ gratefully acknowledges support of the Deutsche Forschungsgemeinschaft (DFG)
through the Transregional Collaborative Research Center SFB/TR7
``Gravitational Wave Astronomy: Methods--Sources--Observation''. 
The work of P.J.\ was also supported in part by the Polish MNiSW grant no.\ N N203 387237.

\end{acknowledgments}


\begin{thebibliography}{99}

\bibitem{LIGO2011}
J.\ Abadie \textit{et al.}
(LIGO Scientific Collaboration and Virgo Collaboration),
Phys.\ Rev.\ D \textbf{83}, 122005 (2011).

\bibitem{Damour2012}
T.\ Damour,
The general relativistic two body problem and the effective one body formalism,
arXiv:1212.3169v1.

\bibitem{BDGNR2011}
L.\ Baiotti, T.\ Damour, B.\ Giacomazzo, A.\ Nagar, and L.\ Rezzolla,
Phys.\ Rev.\ D \textbf{84}, 024017 (2011).

\bibitem{BBLT2012}
E.\ Barausse, A.\ Buonanno, and A.\ Le Tiec,
Phys.\ Rev.\ D \textbf{85}, 064010 (2012).

\bibitem{TBW2012}
A.\ Le Tiec, L.\ Blanchet, and B.\ F.\ Whiting,
Phys.\ Rev.\ D \textbf{85}, 064039 (2012).

\bibitem{JS2012}
P.\ Jaranowski and G.\ Sch\"afer,
Phys.\ Rev.\ D \textbf{86}, 061503(R) (2012).

\bibitem{D2010}
T.\ Damour,
Phys.\ Rev.\ D \textbf{81}, 024017 (2010).

\bibitem{BDTW2010}
L.\ Blanchet, S.\ L.\ Detweiler, A.\ Le Tiec, and B.\ F.\ Whiting,
Phys.\ Rev.\ D \textbf{81}, 084033 (2010).

\bibitem{FS2012}
S.\ Foffa and R.\ Sturani,
The dynamics of the gravitational two-body problem in the post-Newtonian approximation
at quadratic order in Newton's gravitational constant,
arXiv:1206.7087v2.

\bibitem{ADM62}
R.\ Arnowitt, S.\ Deser, and C.\ W.\ Misner,
in \textit{Gravitation: An Introduction to Current Research},
edited by L.\ Witten (John Wiley, New York, 1962), p.\ 227; arXiv:gr-qc/0405109.

\bibitem{DS1991}
T.\ Damour and G.\ Sch\"afer,
Journ.\ Math.\ Phys.\ \textbf{17}, 127 (1991).

\bibitem{DJS2001}
T.\ Damour, P.\ Jaranowski, and G.\ Sch\"afer,
Phys.\ Lett.\ B \textbf{513}, 147 (2001).

\bibitem{BDEF04}
L.\ Blanchet, T.\ Damour, and G.\ Esposito-Far{\`e}se,
Phys.\ Rev.\ D \textbf{69}, 124007 (2004).

\bibitem{FS2011a}
S.\ Foffa and R.\ Sturani,
Phys.\ Rev.\ D \textbf{84}, 044031 (2011).

\bibitem{DJS2008}
T.\ Damour, P.\ Jaranowski, and G.\ Sch\"afer,
in \textit{Proceedings of the 11th Marcel Grossmann Meeting
on Recent Developments in Theoretical and Experimental General Relativity, Gravitation and Relativistic Field Theories},
edited by H.~Kleinert, R.~T.~Jantzen, and R.~Ruffini
(World Scientific, Singapore, 2008),
pp.\ 2490--2492.

\bibitem{Chu2009}
Y.-Z.\ Chu,
Phys.\ Rev.\ D \textbf{79}, 044031 (2009).

\bibitem{J1997}
P.\ Jaranowski,
in \textit{Mathematics of Gravitation. Part II. Gravitational Wave Detection},
edited by A. Kr\'olak, Banach Center Publications
(Institute of Mathematics, Polish Academy of Sciences, Warszawa, 1997),
Vol.\ 41, Part II, pp.\ 55--63.

\bibitem{JS1998}
P.\ Jaranowski and G.\ Sch\"afer,
Phys.\ Rev.\ D \textbf{57}, 7274 (1998);
\textbf{63}, 028802(E) (2001).

\bibitem{DJS2000b}
T.\ Damour, P.\ Jaranowski, and G.\ Sch\"afer,
Phys.\ Rev.\ D \textbf{62}, 044024 (2000).

\bibitem{DJS2000}
T.\ Damour, P.\ Jaranowski, and G.\ Sch\"afer,
Phys.\ Rev.\ D \textbf{62}, 021501(R) (2000).

\bibitem{DJS2000c}
T.\ Damour, P.\ Jaranowski, and G.\ Sch\"afer,
Phys.\ Rev.\ D \textbf{62}, 084011 (2000).

\end{thebibliography}
\end{document}